\documentstyle[preprint,aps]{revtex}
\begin{document}
\title{New enhancements to Feynmans Path Integral for fermions}
\author{Peter Borrmann, Eberhard R. Hilf}
\address{ Department of Physics, University of Oldenburg,
D-26111 Oldenburg, Germany}
%
%    A B S T R A C T
%
\maketitle
\begin{abstract}
We show that the computational effort for the  numerical solution of
fermionic quantum systems, occurring e.g., in quantum chemistry, solid
state physics, field theory  in principle
grows with less than
the square of the particle number for
problems stated in one space
dimension and with less than the cube
of the particle number for problems
stated in three space dimensions. This is proven by representation
of effective algorithms for fermion systems in the framework of the
Feynman Path Integral.
\end{abstract}
\pacs{}
%
%    I N T R O D U C T I O N
%
\narrowtext
\section{Introduction}
When Feynman stated the thermodynamical Path Integral for
fermions (see \cite{fey1} and references therein), he was in doubt
whether this formula could ever be used in numerical applications.
However, in the last few years considerable progress has been made
in the development of Path Integral Monte Carlo algorithms.
Especially D. Ceperly et. al. \cite{ceper} proposed many
very useful refinements.
Nevertheless, the problem of applying
Path Integral methods to fermion systems remains  hard to be solved.
The reasons are the so-called fermion sign
problem and the computational effort, which grows normally proportional
to the fourth power of the number of particles $N$.
In this letter we present a formally exact method with
a computational effort growing only with the cube of the particle
number. \\
We further show, that there are some strong indications, that the
computational effort should in general grow only with the square of
the particle number. For problems stated in one space dimension,
we give an explicit proof for that.
%
%    P A T H I N T E G R A L   M E T H O D
%
\section{Effective Path Integral method}
A quantum $N$ body system in thermal equilibrium can be described
completely by the quantum partition function
\begin{equation}   \label{parfun}
Z = {\rm Tr} \exp\left( -\beta ( {\rm H}_{0} +{\rm H}_{1} ) \right)\; ,
\end{equation}
where H$_{0}$ and H$_{1}$ are the kinetic and potential energy operator.
The Path Integral formalism
leads to a computational convenient equation to calculate Z. Using
Trotter's formula \cite{trott}  Z can be approximated by
\begin{equation} \label{ztrott}
Z = {\rm Tr} \left( \exp(- \frac{\beta}{M} {\rm H}_{0})
\exp(- \frac{\beta}{M} {\rm H}_{1}) \right)^{M} + O(\beta^{3} / M^{2}).
\end{equation}
For a system of polarized fermions in $d$ space dimensions this
leads to the $d N M$ dimensional integral:
\begin{equation} \label{zint}
Z = \left( \frac{1}{N!} \right)^{M}
 \int
\left[ \prod_{\gamma=1}^{M} \prod_{i=1}^{N}
{\rm d}\vec{x}_{i}(\gamma) \right]
 \prod_{\delta=1}^{M} \det A(\delta+1,\delta)
 \exp \left( - \frac{\beta}{M} \sum_{\alpha=1}^{M}
V(\vec{x}_{1}(\alpha), \ldots, \vec{x}_{N}(\alpha)) \right)  \; ,
\end{equation}
with
\begin{equation}  \label{orgmat}
(A(\alpha+1,\alpha))_{k,l} = \left(\frac{M m}
{2 \pi \beta \hbar^{2}} \right)^{d/2}
\exp(- \frac{M m}{2 \beta \hbar^{2}}
 ( \vec{x}_{k}(\alpha+1)-\vec{x}_{l}(\alpha))^{2} ) ,
\end{equation}
and the periodicity condition $\vec{x}_{i}(M+1) = \vec{x}_{i}(1)$. The
particle mass is denoted by $m$.\\
A number of useful modifications to (3), which
improve the convergence vs. the number of timesteps
$M$ (see e.g. \cite{ti1,ti2}), are known.
Because these modifications do not affect any of the following
we will not stress them here.\\
Usually the Metropolis algorithm \cite{met} is adopted to
evaluate (\ref{zint}). The improvements presented in this paper
are based on a careful analysis of the numerical algorithms.\\

Within the Metropolis Monte Carlo procedure a Markov process has to
be generated, under which the probability distribution is stationary.
In our case this condition is fulfilled by the following
widely used procedure, in which the sampling of the probability
distribution is done by generating two different types of random moves
of the particle coordinates.\\
In every microscopic step, all time slices of all particle coordinates
are moved separately through
\begin{equation}
\vec{x}_{i}(\alpha) \rightarrow
\vec{x}_{i}(\alpha) + \Delta \vec{x} \; ,
\end{equation}
where $\Delta \vec{x}$ is a randomly chosen vector.
In every macroscopic step all time slices of every particle coordinate
are moved at once by constant vector.
\begin{equation}
\vec{x}_{i}(\alpha) \rightarrow \vec{x}_{i}(\alpha)
+ \Delta \vec{y}  , \; \; \alpha=1 ... M .
\end{equation}
In both cases a move is accepted, if the absolute value of the ratio
of weight functions, which are simply the integrands in (\ref{zint}),
is greater than a homogeneous random number. Here it is important
that the absolute values have to be taken, because the determinant
occurring in the weight function $W(p)$ becomes sometimes negative.\\
Any observable $X$ can then  be evaluated through
\begin{equation} \label{deR}
\langle X \rangle = \frac{\sum_{p=1}^{G} X(p) {\rm sign} (W(p))}
{\sum_{p=1}^{G} {\rm sign} (W(p))}
\end{equation}
where the summation in (\ref{deR}) runs over all Metropolis steps,
$X(p)$ being the value of X in the p-th step.\\

Albeit the fact, that this method has been applied in some cases
(see e.g. \cite{bh92}), its usefulness is limited by its high
computational costs.
The main effort in the numerical computation of (\ref{zint}) is the
calculation of the determinant. In every complete microscopic motion
the determinant has to be calculated $N \times M$ times.
 The complete algorithm
is therefore of order  $\leq N^{4}$, if standard matrix factorizations ,
which are of order $N^{3}$, are used. The lower sign stands, because
we found that the
number of iterations necessary to achieve a given precision in
the numerical solution decreases with an increasing number of particles.
This results simply from the fact that a larger number of
identical particles yields better statistics.\\
Now observe, that every change of a microscopic coordinate affects only
the i'th row of matrix $A(\alpha+1,\alpha)$ and
the i'th column of matrix $A(\alpha,\alpha-1)$.
This simple fact can be used to reduce the numerical effort by
a factor $n$. The reason for that is that the updating of an
$ L U$ factorization of a matrix after a row or column exchange
can be done by a numerical effort proportional to $N^{2}$.
Descriptions of such algorithms can be found in some
textbooks on matrix computations \cite{lumat}. We thus claim that the
total algorithm is only of order $N^{3}$.\\
Encouraged by the above result and through the fact that the
matrix (\ref{orgmat}) has some hidden symmetries having only
$2 d N$ independent parameters
instead of $N^{2}$, we found
for the special case of systems describable in one space dimension
($d =1 $) a further reduction of the numerical effort.\\
Absorbing the trivial constant factors in (\ref{orgmat}) the
problem is reduced to calculate the determinant of matrices of the
form
\begin{equation}
B_{i,j} = \exp(-\frac{1}{2} (x_{i}-y_{j})^{2}) .
\end{equation}
Using the linear properties of $\det$ we have
\begin{equation}
\det(B) = \det(M)\;
\prod_{j=1}^{N} \exp(-\frac{1}{2}( x_{j}^{2}+y_{j}^{2} )) \;.
\end{equation}
We now show, that the determinant of  matrices of the special form
\begin{equation}   \label{mmat}
 M_{i,j} = \exp( x_{i} y_{j} )
\end{equation}
can be calculated with an effort of only $N^{2}$ operations.
This can most easily be seen using the decomposition
$M = R S T $ of (\ref{mmat}).
\begin{equation}
R = \left(
\begin{array}{ccccc}
1      & y_{1} & y_{1}^{2} &\cdots \\
1      & y_{2} & y_{2}^{2} & \cdots \\
\vdots & \vdots& \vdots    & \vdots \\
1      & y_{N} & y_{N}^{2} & \cdots \\
\end{array}
\right) ,
S = \left(
\begin{array}{ccccc}
0!     &   0   &  0        &   \cdots \\
0      &   1!  &  0        &   \cdots \\
0      &   0   &  2!       &   \cdots \\
\vdots &\vdots & \vdots    &   \ddots \\
\end{array}
\right) ,
T = \left(
\begin{array}{ccccc}
1      & 1     & \cdots     & 1 \\
x_{1}  & x_{2} & \cdots     &  x_{N} \\
x_{1}^2&x_{2}^2& \cdots   &  x_{N}^2  \\
\vdots & \vdots& \vdots & \vdots \\
\end{array}
\right),
\end{equation}
The truncation of these infinite matrices to simple
$(N \times N)$-matrices corresponds to a truncation of
the exponential series at order
$N-1$. For sufficiently large $N$ there is obviously no problem in
doing so and we have
\begin{equation}
\det( M ) = \det(R)  \det(S)  \det(T) .
\end{equation}
The calculation of $\det(S)$ is just trivial and yields a constant for
given $N$. $R$ and $T$ are
Vandermonde matrices, for which the calculation of the determinants
is extremely simple \cite{bellman}. For example for $R$  we have
\begin{equation}           \label{vandermonde}
\det(R) = \prod_{1 \leq j < i \leq N} (x_{i} - x_{j}) .
\end{equation}
The complete calculation of $\det M$
thus involves $N (N-1)$ subtractions
and $N (N-1)$ multiplications altogether. Again using the fact, that
only one coordinate is changed in one microscopic step, the algorithm
to compute the determinant is of order $N$ and the complete Monte Carlo
algorithm thus of order $N^{2}$. \\
Thus we are able to rewrite (\ref{zint}) as
\begin{eqnarray}
Z & =& \left( \frac{1}{N!} \right)^{M}
\left( \frac{M m}{2 \pi \beta \hbar^{2}} \right)^{N M/2}
\left( \prod_{i=1}^{N} \frac{1}{i!} \right)
\int  \prod_{\alpha=1}^{M} \prod_{i=1}^{N} dx_{i}(\alpha)
\exp\left(-\frac{M m}{\beta \hbar^{2}}
\sum_{\alpha=1}^{M} \sum_{j=1}^{N}
x_{j}^{2}(\alpha) \right) \\ \nonumber
&&
\left( \frac{M m}{\beta \hbar^{2} } \right)^{N M}
\left( \prod_{\alpha=1}^{M} \prod_{l<k}^{N}
(x_{k}(\alpha+1)- x_{l}(\alpha))^{2} \right)
\exp \left( - \frac{\beta}{M} \sum_{\alpha=1}^{M}
V(\vec{x}_{1}(\alpha), \ldots, \vec{x}_{N}(\alpha)) \right) \; .
\end{eqnarray}
\section{Conclusion}
Our almost  trivial looking results based on basic linear algebra
contribute a significant  improvement to the Path Integral method.
The decomposition of (\ref{mmat}) found together with the truncation of
the indefinite matrices is somewhat unsatisfactory, because the method
applies in principle only to large particle numbers. Nonetheless our
results should be encouraging to continue
to look for efficient algorithms
to compute the determinant of (\ref{mmat}).\\
Because of the equivalence of the mathematical problem,
quite generally our result implies that any solution method for
quantum systems involving fermions should be bound by a
computational effort of order $N^{2}$.\\
Indeed,  a completely different method
 of computing the Path Integral for
$N$ fermion and boson systems was found, which has this property
\cite{bf,pidft}.
%
%    R E F E R E N C E S
%


\begin{references}
\bibitem{fey1} R.P. Feynman, A.R. Hibbs :
Quantum Mechanics and Path Integrals
New York: McGraw Hill 1965

\bibitem{ceper} D.M. Ceperly,
Phys.Rev.Lett. {\bf 69}, 331 (1992)\\
E.L. Pollock, D.M. Ceperly,
Phys.Rev. {\bf B 30}, 2555 (1984)\\
E.L. Pollock, D.M. Ceperly,
Phys.Rev. {\bf B 36}, 8343 (1987)

\bibitem{trott} M.F. Trotter ,
Proc. Am. Math. Soc. {\bf 10}, 545 (1959)

\bibitem{ti1} M. Takahashi, M. Imada ,
J. Phys. Soc. Jpn. {\bf 53}, 963 (1984)

\bibitem{ti2} M. Takahashi, M. Imada ,
J. Phys. Soc. Jpn. {\bf 53}, 3765 (1984)

\bibitem{met} N. Metropolis, A. Rosenbluth, M.N. Rosenbluth,
A.H. Teller, E. Teller ,
J. Chem. Phys. {\bf 21}, 1087 (1953)

\bibitem{bh92} P. Borrmann, E.R. Hilf,
Z.Phys.{\bf D 26}, S350 (1993)

\bibitem{lumat}
G.H. Golub, C.F. Van Loan, Matrix Computations,
The John Hopkins University Press, Baltimore, Maryland, 1989.\\
J.W. Daniel, W.B. Gragg, L. Kaufmann, G.W. Stewart,
Mathematics of Computation {\bf 30} 772 (1976)\\
P.E. Gill, G.H. Golub, M. Murray, M.A. Saunders,
Mathematics of Computation {\bf 28} 772 (1976)

\bibitem{bellman}  R. Bellmann,
Introduction to Matrix Analysis, 2nd edition,
New Dehli: Tata McGraw-Hill 1979 ( pg. 193 )

\bibitem{bf}  P. Borrmann, G.Franke,
J.Chem.Phys.{\bf 98} 2484 (1993)

\bibitem{pidft} P. Borrmann,
Path Integral Density Functional Theory,
Preprint UOL-THEO3-94-4, Universit{\"a}t Oldenburg, 1994.

\end{references}
\end{document}